\documentclass{osa-article}
\usepackage{cite}
\journal{osajournal}

\begin{document}
	
\title{The synergistic enhancement of spin-phonon interaction in a hybrid system}

\author{Yuan Zhou,\authormark{1,2*} Xin-Ke Li,\authormark{1} Dong-Yan L\"{u},\authormark{1} Yong-Chen Xiong, \authormark{1} Hai-Ming Huang,\authormark{1} and Chang-Sheng Hu\authormark{2\dag}}

\address{\authormark{1}School of Mathematics, Physics and Optoelectronic engineering, Hubei University of Automotive Technology, Shiyan 442002, China\\
\authormark{2}School of Physics, Huazhong University of Science and Technology, Wuhan 430074, China}

\email{\authormark{*}zhouyuan@huat.edu.cn\\\authormark{\dag}hucs908@foxmail.com} 

\homepage{https://orcid.org/0000-0003-3356-1800} 


\begin{abstract}
	
The investigation on significantly enhancing the coupling to
NV centers at single-quanta level is of great key point
to further explore its application in quantum information processing (QIP).
We here study a joint scheme to further enhance NV-phonon
coherent coupling with two methods working together in a hybrid
optomechanical systems.
Both methods are mechanic-induced mode field coupling (MFC) leading to the modification of the spatial distribution of the optical field,
and the mechanical parametric amplification (MPA) realized by modulating the mechanical spring constant in time, respectively.
With the joint assistance of MFC and MPA,
the coherent coupling between the NV spin and one supermode of the mechanical resonators (MRs) can be further enhanced significantly,
with the rate $\propto \bar{n}_{\text{cav}}e^{r}$, several potential applications on this proposal are also discussed in this work.
For the ultimate target of enhancing the coupling to NV spin at single-quanta level,
this attempt may provide a promising spin-phonon platform for implementing the more active control.
\end{abstract}

\section{Introduction}

Implementing a controllable and enough strong coupling to a quantum unit at single-quanta
level is a most desirable basic goal in quantum information processing (QIP)
\cite{PhysRevLett.83.4204,PhysRevX.5.031031,PhysRevLett.110.213604,PhysRevLett.95.087001,NRMAtature,NCYaoNY,PhysRevX.4.031022}.
Basing on this coherent coupling, first we can carry out a complete and fast control
to the qubits directly or indirectly at single-quanta level \cite{NatLadd,Science326.108.111},
which underlies the applications of the quantum simulation \cite{RevModPhys.86.153,NPhyCJM},
manipulation \cite{RevModPhys.85.623,PhysRevX.4.021052}, and metrology \cite{PhysRevX.10.031003}.
Secondly, we can also explore many interesting and essential physics \cite{Mitchell:19}, 
such as single photon or phonon technology \cite{RevModPhys.87.1379,PhysRevLett.107.063602,Mitchell:16},
chiral quantum science \cite{nature21037,Immo2015,Zhou:21}, etc.

Working as a point defect in diamond, the nitrogen-vacancy
(NV) center integrated in a hybrid quantum system has
recently emerged as one of the leading candidates for QIP
thanks to their excellent spin properties
\cite{Doherty2013,Bar2013,P2010Quantum},
such as solid-state spins with atom-like properties and without additional trap device
\cite{Zhu2011Coherent,PhysRevX.1.011007},
precise implantation and easy scalability \cite{PhysRevLett.105.140502,PhysRevLett.105.210501},
longer coherence times even at ambient conditions \cite{Bar2013,P2010Quantum,PhysRevLett.113.237601},
and the convenient preparation, manipulation, and readout of its quantum state,
etc \cite{PhysRevLett.125.153602,PhysRevLett.117.015502}.
Utilizing NV spins in hybrid systems,
significant theoretical and experimental investigations
have been carried out to realize quantum simulation and quantum state manipulating
\cite{PhysRevLett.110.156402,PhysRevLett.116.143602,PhysRevLett.113.020503,PhysRevApplied.4.044003,
PhysRevA.83.022302,Zhou_2021,PhysRevA.98.052346,PhysRevA.96.062333}.
In recent years, more and more attentions are devoted to
the applications of NV centers in quantum acoustics area,
which also leads to a growing interest in studying and
exploiting the coherent spin-phonon coupling
\cite{PhysRevX.5.031031,PhysRevX.6.041060,PhysRevLett.112.116403}.
However, it is still a huge challenge for us to significantly
enhance the spin-phonon coupling at single-quanta level
by the means available \cite{PhysRevLett.125.153602}.

In this work, we present a combined scheme to enhance the spin-phonon coupling in a hybrid setup,
which is consist of a single NV spin and three optical cavities dispersively coupled with three mechanical resonators (MRs)
\cite{RevModPhys.86.1391}.
To further enhance the spin-phonon coupling in this spin-cavity-resonator tripartite system, there are mainly two key points in our proposal.
First we can modify the spatial distribution of electric field $\vec{E}(x,y,z)$ in the cavity through the mechanical displacement,
which is named as the mode field coupling (MFC) \cite{PhysRevLett.118.133603}.
Importantly, the spin-phonon interaction can be controlled and enhanced by the optical field intensity
with the rate $\Lambda\propto\vec{E}(x,y,z)\sim\bar{n}_{\text{cav}}$,
resulting in optically controlled spin-phonon coherent manipulation.
Secondly, we meanwhile apply the mechanical parametric amplification (MPA)
to the MR through modulating its spring constant in time \cite{PhysRevLett.67.699,PhysRevA.64.063803,PhysRevA.66.025801,PhysRevLett.114.093602,
PhysRevLett.107.213603,NCLemonde,Liao_2014,Szorkovszky_2014,PhysRevLett.120.093602,PhysRevLett.122.030501,
PhysRevLett.125.153602,PhysRevLett.120.093601}.
In the squeezed frame, we can further enhance the spin-phonon coupling
with an exponential rate $\Lambda\propto e^{r}$ in this tripartite system \cite{PhysRevLett.125.153602,Burd2021}.
In a word, taking advantage of this joint assistance of MFC and MPA,
we have achieved the goal of further strengthening the coherent
spin-phonon coupling at single-quanta level,
comparing with the previous investigations on strengthening the spin-phonon coupling.
Besides we also have discussed several potential applications basing on this tripartite interaction system.
We stress that this scheme may provide a promising phonon-mediated platform for implementing the more active control to NV spins.

\section{The scheme}

\begin{figure}[t]
\centering\includegraphics[width=10cm]{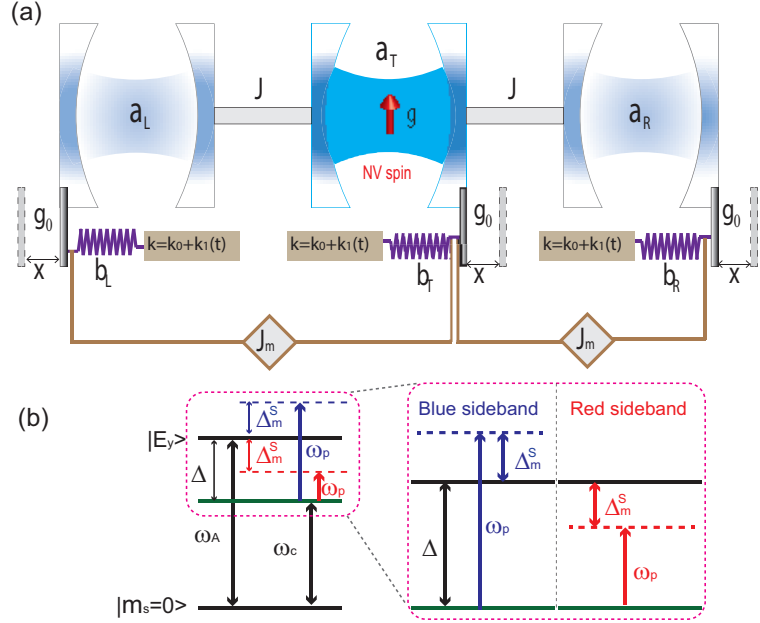}
\caption{(Color online) The schematic of our hybrid system.
(a) Three identical optical cavities with frequency $\omega_{c}$ are arranged in a row,
the central cavity couples to the left cavity and the right cavity with the identical coupling strength $J$, through exchanging photons via optical fibres.
Each cavity dispersively couples to the corresponding MR with the coupling strength $g_{0}$. The additional second-order nonlinear pump is applied to each MR,
which can be realized by modulating the spring constant in time.
The central mechanical resonator is additionally coupled to the other two bilateral MR with the same coupling rate $J_{m}$.
A single NV center is placed inside the central cavity and interacts with this cavity mode with the coupling strength $g$.
(b) Energy level diagram illustrating the blue and red sideband transitions for the tripartite interaction quantum system.}
\end{figure}

We have design this hybrid setup as illustrated in Fig.~1(a),
three identical optical cavities with frequency $\omega_{c}$ are arranged one by one,
and these optical modes are named as modes $\hat{a}_{L}$, $\hat{a}_{T}$,
and $\hat{a}_{R}$, respectively.
The central cavity is symmetrically connected to the two bilateral cavity with the identical optical fibers.
As thus, through the exchange-photon process, the central cavity mode $\hat{a}_{T}$ will interact with the two bilateral cavities with the same coupling rate $J$.
Each cavity also dispersively couples to a identical mechanical resonator (MR) with the same coupling rate $g_{0}$.
For these three identical MRs, the fundamental frequency are all $\omega_{m}$, and these mechanical modes are correspondingly named as $\hat{b}_{L}$, $\hat{b}_{T}$, and $\hat{b}_{R}$.
Moreover, we add an additional second-order nonlinear pump on each resonator,
which can be implemented easily through modulating the mechanical spring constant in time.
The central MR $\hat{b}_{T}$ additionally couples to the other two bilateral MRs $\hat{b}_{L}$ and $\hat{b}_{R}$ with the same coupling rate $J_{m}$.

Beside, as illustrated in Fig.~1(a),
a single NV center is placed inside the central cavity.
In the optical frequency domain, the optical mode $\hat{a}_{T}$
will induce the NV spin's quantum transition
between the excited state $|E_{y}\rangle$ and the ground state $|m_{s}=0\rangle$
with the coupling rate $g$.
The energy level structure of a single NV center is shown in Fig. 1(b).
The ground state and the excited state are denoted as $|m_{s}=0\rangle\equiv|0\rangle$ and $|E_{y}\rangle\equiv|1\rangle$,
the optical transition frequency between them is $\omega_{A}\sim 2\pi\times 470$ THz.
For the single NV center, this two-level system $\{|0\rangle,|1\rangle\}$ can be considered as a spin-1/2
particle with the Pauli matrix definitions $\hat{\sigma}_{z}\equiv(|1\rangle\langle1|-|0\rangle\langle0|)/2$,
$\hat{\sigma}_{+}\equiv|1\rangle\langle0|$, and $\hat{\sigma}_{-}\equiv|0\rangle\langle1|$.

Therefore, according to the Appendix 8.1-8.3
we can get the total Hamiltonian for describing this hybrid system $(\hbar =1)$
\begin{equation}
\hat{H}_{\text{Total}}=\hat{H}_{1}+\hat{H}_{2}+\hat{H}_{3},
\end{equation}
where
\begin{eqnarray}
\hat{H}_{1}=\sum_{j=L,R,T}[\Delta_{m}\hat{b}_{j}^{\dag}\hat{b}_{j}
-\frac{\Omega_{p}}{2}(\hat{b}_{j}^{2}+\hat{b}_{j}^{\dag 2})]
+J_{m}\hat{b}_{T}^{\dag}(\hat{b}_{L}+\hat{b}_{R})+h.c.,\notag
\end{eqnarray}
\begin{eqnarray}
\hat{H}_{2}=\sum_{j=L,R,T}\omega_{c}\hat{a}_{j}^{\dag}\hat{a}_{j}+\omega_{A}\hat{\sigma}_{z}
+g\hat{a}_{T}^{\dag}\hat{\sigma}_{-}+J\hat{a}_{T}^{\dag}(\hat{a}_{L}+\hat{a}_{R})+h.c.,\notag
\end{eqnarray}
\begin{eqnarray}
\hat{H}_{3}=\sum_{j=L,R,T}[-g_{0}\hat{a}_{j}^{\dag}\hat{a}_{j}(\hat{b}_{j}^{\dag}\exp{i\omega_{p}t}+h.c.)].\notag
\end{eqnarray}
In equation (1), under the rotating frame with frequency $\omega_{p}$, the first item $\hat{H}_{1}$ is the Hamiltonian for describing these three MRs with the second-order nonlinear interaction, including their pairwise interactions between the central mode $\hat{b}_{T}$ and the bilateral modes $\hat{b}_{R,L}$.
The second item $\hat{H}_{2}$ describes the NV spin and optical cavities, with the spin-cavity interaction and the pairwise interactions between the central mode $\hat{a}_{T}$ and the bilateral modes $\hat{a}_{R,L}$.
The last item $\hat{H}_{3}$ means the Hamiltonian for describing the dispersive interactions between the cavities and the corresponding mechanical resonators.

According to the Appendix 8.4, we can simplified the total Hamiltonian for this system,
and obtain an effective Hamiltonian with the tripartite interactions (spin-photon-phonon).
\begin{eqnarray}
\hat{H}_{\text{eff}}\approx\Delta\hat{\sigma}_{z}+\Delta_{m}^{S}\hat{b}_{0}^{\dag}\hat{b}_{0}
+\frac{gg_{0}^{S}}{2J}(\hat{b}_{0}+\hat{b}_{0}^{\dag})(\hat{a}_{0}^{\dag}\hat{\sigma}_{-}+h.c.).
\end{eqnarray}

\section{Enhancing the spin-phonon coupling}
\begin{figure}[t]
	\centering\includegraphics[width=12cm]{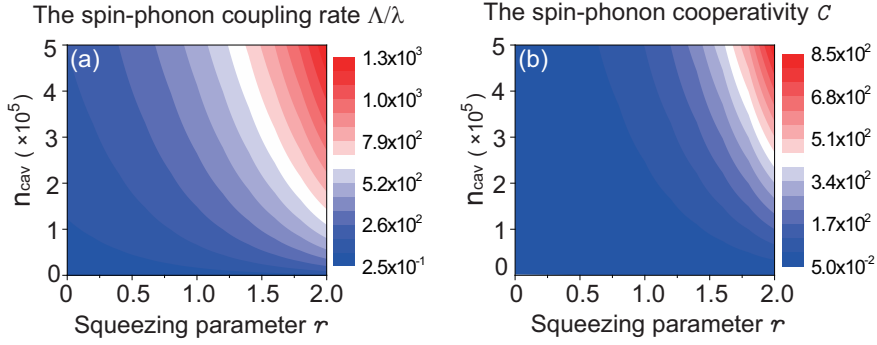}
	\caption{(Color online)
		(a) The spin-phonon coupling enhancement $\Lambda/\lambda$ and
		(b) the cooperativity enhancement $C$ versus the squeezing
		parameter $r$ and the photon number $n_{\text{cav}}$
		of the cavity mode $\hat{a}_{0}$, with $g_{0}=0.001g$, $J=10g$, $\lambda/2\pi=0.1$MHz, $g/2\pi=1$GHz, the effective mechanical dissipation $\Gamma_{m}^{S}/2\pi\sim1$MHz,
		and the NV spin decay rate $\gamma/2\pi\sim15$MHz.  }
\end{figure}

\begin{figure}[t]
	\centering\includegraphics[width=12cm]{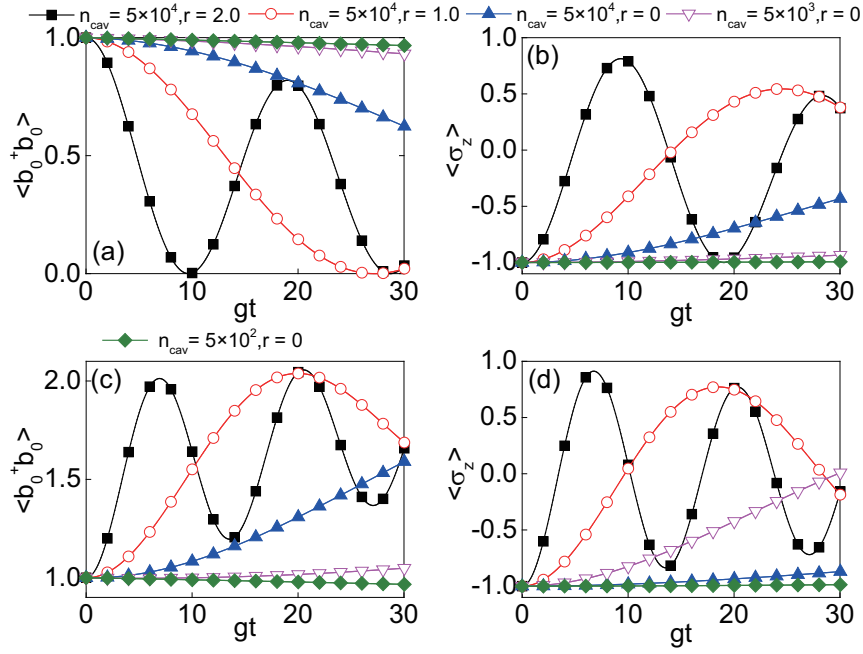}
	\caption{(Color online) The dynamical population of the phonon number $\hat{b}_{0}^{\dag}\hat{b}_{0}$ and the spin operator $\hat{\sigma}_{z}$ according to the J-C model ((a) and (b)) and the anti J-C model ((c) and (d)), with different $n_{\text{cav}}$ and $r$. The parameters are $g_{0}=0.001g$, $J=10g$, the effective mechanical dissipation $\Gamma_{m}^{S}\sim0.001g$,
		and the NV spin decay rate $\gamma\sim 0.02g$. This system is initially prepared in state $|\phi(0)\rangle=|1\rangle_{m}|0\rangle_{s}$. }
\end{figure}

We consider the cavity is pumped with a large coherent field with an average photon number $\overline{n}_{\text{cav}}\equiv\sqrt{n_{\text{cav}}}$.
Therefore we can write the cavity field as $\hat{a}_{0}=\overline{n}_{\text{cav}}+\delta\hat{a}_{0}$.
Neglecting the quantum fluctuations $\delta\hat{a}_{0}$ (valid for $n_{\text{cav}}\gg1$),
we can acquire the effective Rabi type Hamiltonian
\begin{eqnarray}
\hat{H}_{\text{eff}}\simeq\Delta\hat{\sigma}_{z}+\Delta_{m}^{S}\hat{b}_{0}^{\dag}\hat{b}_{0}
+\frac{\overline{n}_{\text{cav}}gg_{0}^{S}}{2J}(\hat{b}_{0}+\hat{b}_{0}^{\dag})\hat{\sigma}_{x}.
\end{eqnarray}
Next, in the interaction picture (IP),
we transfer Eq.~(3) into an equivalent expression with the relations $g_{0}^{S}=g_{0}e^{r}\cos \omega_{p}t$ and $\omega_{p}=\Delta-\Delta_{m}^{S}$ (the red sideband detuning),
through discarding the high frequency oscillation terms.
\begin{eqnarray}
\hat{H}_{\text{JC}}^{\text{IP}}\simeq\Lambda (\hat{b}_{0}\hat{\sigma}_{+}+\hat{b}_{0}^{\dag}\hat{\sigma}_{-}),
\end{eqnarray}
and this is the Jaynes-Cumming (J-C) type Hamiltonian.
On the contrary, when we assume $\omega_{p}=\Delta+\Delta_{m}^{S}$ (the blue sideband detuning), we can also achieve the anti J-C model,
\begin{eqnarray}
\hat{H}_{\text{A-JC}}^{\text{IP}}\simeq\Lambda (\hat{b}_{0}\hat{\sigma}_{-}+\hat{b}_{0}^{\dag}\hat{\sigma}_{+}).
\end{eqnarray}
Whether it is J-C model or anti J-C model,
we can get an enhanced coupling strength with the effective coupling strength
\begin{eqnarray}
\Lambda\equiv\frac{\overline{n}_{\text{cav}}gg_{0}e^{r}}{4J}.
\end{eqnarray}

For single NV center, we can achieve a traditional weak spin-phonon coupling at single quantum level with the strength $\lambda/2\pi\leq 0.1$ MHz.
To quantify the enhancement of the spin-phonon coupling, we exploit the cooperativity $C=\Lambda^{2}/\Gamma_{m}\gamma$.
Here, $\Gamma_m=n_\text{th}\kappa_{m}$ and $\gamma$ correspond to the effective mechanical dissipation rates and the decay rate of the spin, respectively.
Note that in presence of the mechanical amplification,
the noise coming from the mechanical bath is also amplified.
To circumvent this detrimental effect, a possible strategy is to use the dissipative squeezing approach to keep the mechanical mode in its ground state in the squeezed frame \cite{PhysRevLett.114.093602,NCLemonde,PhysRevA.88.063833}.
This steady-state technique has already been implemented experimentally \cite{Wollman952}.
In this case, we can obtain the engineered effective dissipation rate $\Gamma_{m}^{S}$ in the squeezed frame.
Therefore, we can also define the effective cooperativity $C=\Lambda^{2}/\Gamma_{m}^{S}\gamma$.

In Fig.~2, we plot the spin-phonon coupling enhancement $\Lambda/\lambda$ and
the cooperativity enhancement $C$, versus the squeezing parameter $r$ and photon number $n_{\text{cav}}$ of the classical driven field on the mode $\hat{a}_{0}$. We find that as we increase the squeezing parameter $r$ and the photon number $n_{\text{cav}}$,
we can achieve a distinct enhancement in the spin-phonon coupling, thus directly giving rise to the cooperativity enhancement.

Furthermore, in Fig.~3, we also plot the dynamical population of the phonon number operator $\hat{b}_{0}^{\dag}\hat{b}_{0}$ and the spin operator $\hat{\sigma}_{z}$ according to the J-C model (in Eq.~(4)) and the anti J-C model (in Eq.~(5)), with the different parameters, such as $n_{\text{cav}}$ and $r$.
The numerical results above show the distinct quantum dynamics of this spin-phonon system for different cases,
in which, the spring constant is modulated or not, and $n_{\text{cav}}$ is increased from $5\times10^2$ to $5\times10^4$.
Therefore, with this joint assistance of the mechanical squeezing (with parameter $r$) and the classical driving of the mode $\hat{a}_{0}$ (with intensity $n_{\text{cav}}$), the system can be pumped and driven from the weak-coupling regime to the strong-coupling, or even to the ultrastrong-coupling regime.

\section{The enhancement of photon-spin-phonon interaction}
\begin{figure}[t]
	\centering\includegraphics[width=12cm]{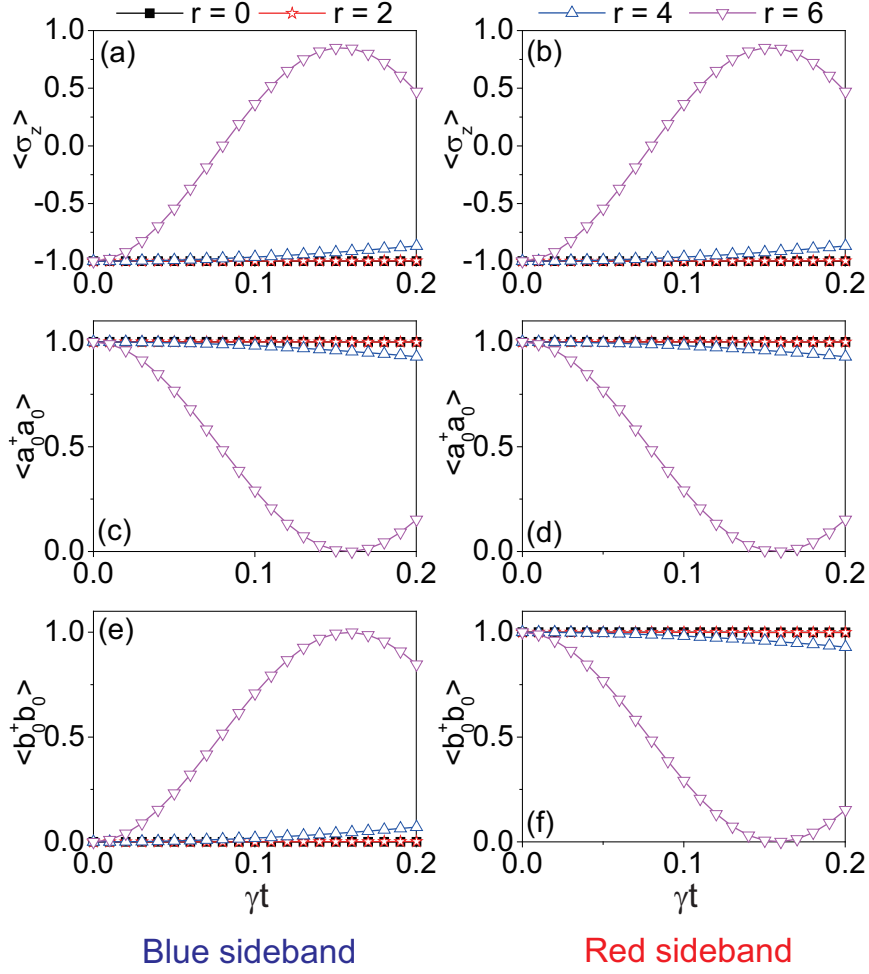}
	\caption{(Color online) The dynamical population of the spin operator $\hat{\sigma}_{z}$ (a) and (b), (c) and (d) the number operators of the optical mode $\hat{a}_{0}^{\dag}\hat{a}_{0}$, (e) and (f) the phonon mode $\hat{b}_{0}^{\dag}\hat{b}_{0}$. In which, (a) (c) (e) correspond to the blue-sideband condition, and (b) (d) (f) corresponds to the red-sideband condition, with different squeezing parameter $r$. The parameters are $g_{0}\sim\gamma$, $J=2.8\times 10^{3}\gamma$, $g\sim70\gamma$,
		the effective mechanical dissipation and the cavity decay rate are assumed $\Gamma_{m}^{S}\sim0.001\gamma$, $\kappa\sim0.1\gamma$,
		and the NV spin decay rate $\gamma/2\pi\sim 15$ MHz. This tripartite system is initially prepared in states  $|\psi(0)\rangle=|1\rangle_{o}|0\rangle_{m}|0\rangle_{s}$ (blue sideband) and $|\psi(0)\rangle=|1\rangle_{o}|1\rangle_{m}|0\rangle_{s}$ (red sideband), respectively.}
\end{figure}
On the other hand, if $\overline{n}_{\text{cav}}$ is too weak,
and the quantum fluctuations $\delta\hat{a}_{0}$ will dominate the supermode $\hat{a}_{0}$.
As a result, we can get the effective tripartite interaction Hamiltonian from Eq.(2)
with two different kinds of expression in IP.
For the first condition, such as the blue sideband,
when the resonance condition satisfies $\omega_{p}=\Delta+\Delta_{m}^{S}$,
we can also discard the high frequency oscillation terms,
and get the blue sideband effective three quantum system Hamiltonian,
\begin{eqnarray}
\hat{H}_{\text{Blue}}^{\text{IP}}\simeq\Lambda_{0}
(\hat{\sigma}_{-}\hat{b}_{0}\hat{a}_{0}^{\dag}+\hat{\sigma}_{+}\hat{b}_{0}^{\dag}\hat{a}_{0}).
\end{eqnarray}
For the second case, such as the red sideband,
when the resonance condition satisfies $\omega_{p}=\Delta-\Delta_{m}^{S}$,
discarding the high frequency oscillation terms, we can get
\begin{eqnarray}
\hat{H}_{\text{Red}}^{\text{IP}}\simeq\Lambda_{0}
(\hat{\sigma}_{-}\hat{b}_{0}^{\dag}\hat{a}_{0}^{\dag}+\hat{\sigma}_{+}\hat{b}_{0}\hat{a}_{0}).
\end{eqnarray}
In this tripartite interaction quantum system,
the effective coupling strength is
\begin{eqnarray}
\Lambda_{0}=\frac{gg_{0}e^{r}}{4J}.
\end{eqnarray}

In Fig.~4, we make the simulations on this tripartite interaction system according to Eq.~(7) and Eq.~(8), and then plot the dynamical population of the phonon number operator $\hat{b}_{0}^{\dag}\hat{b}_{0}$, the photon number operator $\hat{a}_{0}^{\dag}\hat{a}_{0}$,
and the spin operator $\hat{\sigma}_{z}$ with the different squeezing parameter $r$ .
The numerical results above, evidently show us that we can strength this tripartite
interaction through increasing the squeezing parameter $r$.

\section{The application on this proposal}

\subsection{Entangling collective NV spins dynamically}
\begin{figure}[t]
	\centering\includegraphics[width=12cm]{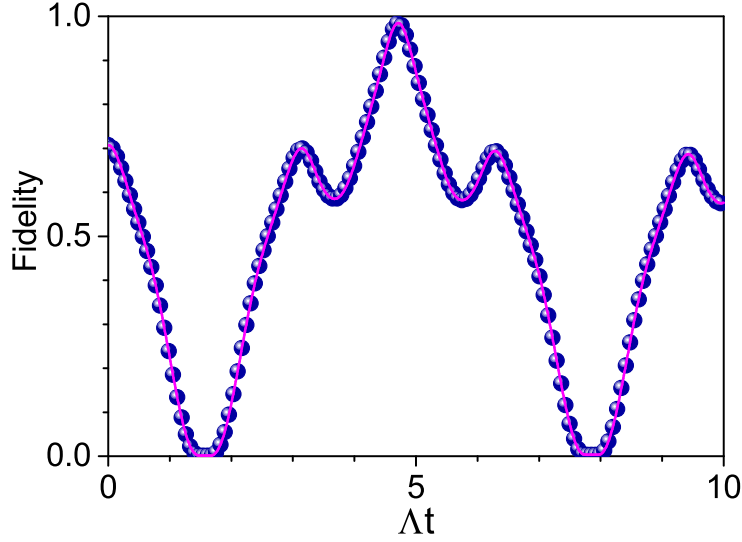}
	\caption{(Color online) The dynamical fidelity of the target entangled GHZ state for four NV spins.
		In which, the initial state is $|\psi_{\text{system}}(0)\rangle=|0\rangle_{m}|0000\rangle_{s}$,
		and the target GHZ state is $|\psi_{\tau}^{\text{NV}}\rangle=[ e^{-i\pi/4}|0000\rangle_{s}+e^{i\pi/4}|1111\rangle_{s}]/\sqrt{2}$.
		The parameters are assumed as the squeezing parameter $r\simeq 4.0$, $g_{0}\sim 0.001g$, $J\sim10g$, $g/2\pi\sim 1.0$ GHz, $\overline{n}_{\text{cav}}\sim 10^{4}$, the effective mechanical dissipation $\Gamma_{m}^{S}\sim0.001\gamma$, and the NV spin decay rate $\gamma/2\pi\sim 15$ MHz. }
\end{figure}
In this section, the first type of potential application for this proposal is
that we can entangle the separated NV spins.
We assume a certain number of NV spins are set separately in this central optical cavity.
According to the Eq.~(3), we can obtain the effective Hamiltonian as follow,
\begin{eqnarray}
\hat{H}_{\text{eff}}\simeq\Delta_{m}^{S}\hat{b}_{0}^{\dag}\hat{b}_{0}
+\sum_{k}\Lambda^{k}(\hat{b}_{0}+\hat{b}_{0}^{\dag})\hat{\sigma}_{x}^{k}.
\end{eqnarray}
Here, we assume $\Delta=\omega_{A}-\omega_{c}=0$,
and the effective coupling to the $k$th NV spin is $\Lambda^{k}=\overline{n}_{\text{cav}}^{k}g^{k}g_{0}e^{r}/4J$.
We stress that this inhomogeneous coupling strength is mainly caused by the differences of location of NV spins in the cavity,
which maps to the factors $\overline{n}_{\text{cav}}^{k}$ and $g^{k}$.
In this scheme, one can reduce this system disorder through implanting NV spins precisely with the advanced processing techniques.
By discarding this weak adverse effect, we can rewrite the effective Hamiltonian in the interaction picture (IP) as
\begin{eqnarray}
\hat{H}_{\text{eff}}^{\text{IP}}\simeq\Lambda(\hat{b}_{0}e^{-i\Delta_{m}^{S}t}+\hat{b}_{0}^{\dag}e^{i\Delta_{m}^{S}t})\hat{J}_{x}.
\end{eqnarray}
In which, the coupling is identical $\Lambda^{k}\equiv\Lambda$,
and so we can utilize the collective spin operator in above equation with the definition $\hat{J}_{x}\equiv\sum_{k}\hat{\sigma}_{x}^{k}$.
We note this type interaction corresponds to the so-called M{\o}lmer-S\o{}rensen (MS) gate \cite{PhysRevLett.82.1971,PhysRevA.62.022311},
which is utilized for generating the multiparticle entanglement.
And its dynamics of this system is governed by the unitary evolution operator $\hat{U}_{\text{IP}}(t)=\exp(-i\hat{H}_{\text{eff}}^{\text{IP}}t)$.
Taking advantage of the Mangnus formula \cite{Takahashi_2017},
we can get $\hat{U}_{\text{IP}}(\tau)\simeq\exp(-i\Lambda^{2}\hat{J}_{x}^{2}\tau/\Delta_{m}^{S})$
when $\tau=2 n \pi /\Delta_{m}^{S}$ for the integer number $n$. This means that the mechanical mode is decoupled from the NV spins at this moment.

Note that as this operator has no contribution from the mechanical modes, thus
in this instance the system gets insensitive to the states of the mechanical modes.
Starting from the initial state of the mechanical mode and NV spins $|\psi_{\text{system}}(0)\rangle=|0\rangle_{m}|00 \cdots 00\rangle_{s}$,  we can obtain the target entangled state of the collective NV spins with the form $|\psi_{\tau}^{\text{NV}}\rangle=[ e^{-i\pi/4}|00 \cdots 00\rangle_{s}+e^{i\pi/4}(-1)^{N}|11 \cdots 11\rangle_{s}]/\sqrt{2}$, which is the well-known Greenberger-Horne-Zeilinger (GHZ) type state with $N$ the number of the spins.
Then we plot this numerical simulation result in Fig.~5.
As illustrated in Fig.~5, taking the realistic condition such as the NV
decay rate and mechanical dissipation all into our considerations,
we can quickly entangle NV spins with a high fidelity of more than $0.98$ in this scheme.

\subsection{Local cooling one supermode of triple resonators with an NV ensemble}
\begin{figure}[t]
	\centering\includegraphics[width=12cm]{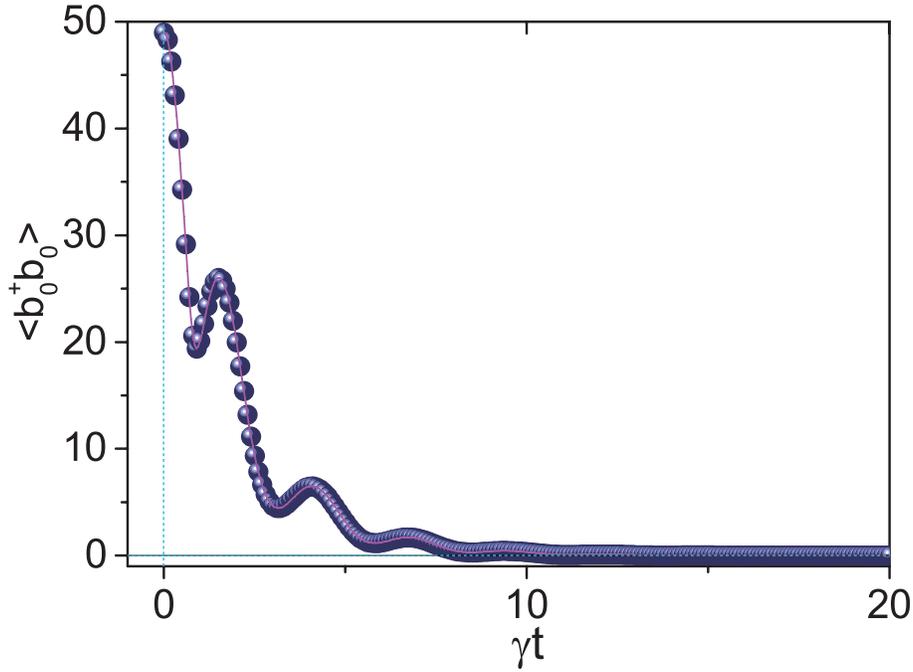}
	\caption{(Color online) The dynamical population of this mechanical supermode $\hat{b}_{0}$ with the assumption of its initial average phonon number $\langle\hat{b}_{0}^{\dag}\hat{b}_{0}\rangle\simeq50$. In which, the following parameter are the squeezing parameter $r\simeq 2.0$, $\overline{n}_{\text{cav}}\sim 100$, $g_{0}\sim 0.001g$, $g\sim 66\gamma$, $J\sim10g$, the effective mechanical dissipation $\Gamma_{m}^{S}\sim0.001\gamma$, and the NV spin decay rate $\gamma/2\pi\sim15$ MHz, respectively. }
\end{figure}
On the other hand, we stress that another potential application on this scheme is to cool down the mechanical supermode to its ground state efficiently with the NV center ensemble (NVE). 
Here, we assume a number of NV centers are set inside the central optical cavity, which form an NVE.
Taking the Eq.~(4) and Eq.~(10) into our considerations, we can obtain the effective Hamiltonian in IP for this hybrid system,
\begin{eqnarray}
\hat{H}_{\text{eff}}^{\text{IP}}\simeq\sum_{k=1}^{N}\Lambda^{k} (\hat{b}_{0}\hat{\sigma}_{+}^{k}+\hat{b}_{0}^{\dag}\hat{\sigma}_{-}^{k}).
\end{eqnarray}
Similarly, we can also ignore this weak system-disorder adverse effect according to the advanced processing techniques.
Then we can rewrite this effective Hamiltonian as
\begin{eqnarray}
\hat{H}_{\text{eff}}^{\text{IP}}\simeq \Lambda (\hat{b}_{0}\hat{J}_{+}+\hat{b}_{0}^{\dag}\hat{J}_{-}),
\end{eqnarray}
with the collective spin operator $\hat{J}_{\pm}\equiv\sum_{k=1}^{N}\hat{\sigma}_{\pm}^{k}$.
In the condition of weak excitations and $N\gg1$,  we can map
the collective spin operators $\hat{J}_{\mp}$ into the boson operators
$\hat{d}$ and $\hat{d}^{\dagger}$ in the Holstein-Primakoff representation,
with $\hat{J}_{+}\simeq\sqrt{N}\hat{d}^{\dagger}$, $\hat{J}_{-}\simeq\sqrt{N}\hat{d}$,
and $\hat{J}_{z}\simeq(\hat{d}^{\dagger}\hat{d}-\frac{N}{2})$.
Then we can carry out this goal of cooling down one mechanical supermode $\hat{b}_{0}\equiv(\hat{b}_{L}^{S}-\hat{b}_{R}^{S})/\sqrt{2}$ to its ground state efficiently.
As illustrated in Fig.~(6), we can transform this mechanical mode into its quantum ground state completely at the time of $\sim\frac{10}{\gamma}$ through numerically solving the quantum master equation.
This result indicates that we can implement this cooling process successfully in this setup.

\section{The experimental parameters }
To examine the feasibility of our scheme in realistic experiment,
we now discuss the relevant experimental parameters.
We consider a high quality optical cavity with frequency $\omega_{c}/2\pi\sim 470$ THz and $Q\sim10^{6}-10^{8}$,
and we can assume its coupling strength to single NV center can reach to $g/2\pi\sim10$ GHz \cite{Jiang2020,Wang:18,Wu:18,Fang:19,PhysRevLett.109.033604,RevModPhys.87.347,doi:10.1063/1.4904909}.
While for the mechanical resonator with frequency $\omega_{m}/2\pi\sim 1-10$ GHz and $Q\sim10^{5}-10^{6}$,
the optic-mechanic coupling to the cavity mode will be $g_{0}/2\pi\sim1-10$ MHz \cite{RevModPhys.86.1391}.
For single NV center, the lifetime of its excited state is about $10$ ns,
so its spontaneous decay rate of the excited state is about $\gamma/2\pi\sim15$ MHz 
\cite{Khalid2015,PhysRevB.95.205420,Astner2018}.
Considering the dynamical process for entangling the NV spins in this work,
we can obtain the GHZ state at the time of $\sim0.35$ ns.
In compared with this time interval, we think its coherence time is enough for implementing this scheme.

\section{Conclusion}

In summary, we propose a protocol to further enhance the spin-phonon coupling at single-quanta level
with two methods joint working together,
which are the MFC and MPA respectively.
Importantly, in our scheme, we can enhance the coherent spin-phonon interaction
not only by the optical field intensity with rate $\sim\bar{n}_{\text{cav}}$,
but also by the amplified zero-field fluctuation of the mechanical mode
with rate $\sim e^{r}$.
In a word, taking advantage of the joint assistance of both amplification means,
we can realize the goal of further strengthening the coherent
spin-phonon coupling at single-quanta level in this hybrid system.
Besides we also have make a brief discussion on the potential applications of this tripartite interaction system.
We stress that this investigation may provide us a more promising direction for implementing the active control of the spin-phonon coupling at single quanta-level.

\section{Appendix}

\subsection{The Hamiltonian of mechanical modes with second-order nonlinear interaction}

The Hamiltonian for the $j$th mechanical system with a modulated spring constant can be expressed as
\begin{eqnarray}
\hat{H}_{mj}=\frac{\hat{p}_{j}^{2}}{2M_{j}}+\frac{1}{2}k_{0}\hat{x}_{j}^2+\frac{1}{2}k_{1}(t)\hat{x}_{j}^2,
\end{eqnarray}
where $j=\{R,T,L\}$, and $M_{j}$ is the effective mass of the $j$th mechanical resonator.
Expressing the momentum operator $\hat{p}_{j}$ and the displacement operator $\hat{x}_{j}$ with the oscillator operator $\hat{a}_{j}$ of the fundamental oscillating mode and
the zero field fluctuation $x_{\text{zpf}}=\sqrt{\hbar/2M_{j}\omega_{m}}$, i.e., $\hat{p}_{j}=-i(M_{j}\hbar\omega_m/2)^{1/2}(\hat{b}_{j}-\hat{b}^{\dag}_{j})$ and $\hat{x}_{j}=x_{\text{zpf}}(\hat{b}^{\dag}_{j}+\hat{b}_{j})$,
we obtain ($\hbar =1$),
\begin{eqnarray}
\hat{H}_{mj}=\omega_{m}\hat{b}_{j}^{\dag}\hat{b}_{j}-\Omega_{p}\cos2\omega_{p}t(\hat{b}_{j}+\hat{b}_{j}^{\dag})^{2}.
\end{eqnarray}
Here we assume that these are three identical mechanical resonators,
with the intrinsic frequency $\omega_{m}=\sqrt{k_{0}/M_{j}}$,
the time-dependent spring constant $k_{1}(t)=\delta k\cos2\omega_{p}t$,
and the nonlinear coefficient $-\delta k{x_{zf}}^{2}/2\equiv\Omega_{p}$.
Utilizing the frame rotating with frequency $\omega_{p}$ and dropping the terms ($\sim \hat{b}_{j}^{\dag}\hat{b}_{j}$) that explicitly oscillate in time,
then we can acquire the Hamiltonian with the second-order nonlinear interaction for the $j$th mechanical resonator
\begin{eqnarray}
\hat{H}_{mj}=\Delta_{m}\hat{b}_{j}^{\dag}\hat{b}_{j}-
\frac{\Omega_{p}}{2}(\hat{b}_{j}^{2}+\hat{b}_{j}^{\dag 2}),
\end{eqnarray}
where $\Delta_{m}=\omega_{m}-\omega_{p}$.
For simplicity, we have assume that $\Omega_{p}$ is a real parameter.

Furthermore, in this scheme, the central mechanical mode $\hat{b}_{T}$
also interacts with another two bilateral mechanical modes $\hat{b}_{L}$ and $\hat{b}_{R}$.
Assuming the identical coupling strength as $J_{m}$,
we can obtain their Hamiltonian as follow
\begin{eqnarray}
\hat{H}_{1}=\sum_{j=L,R,T}[\Delta_{m}\hat{b}_{j}^{\dag}\hat{b}_{j}
-\frac{\Omega_{p}}{2}(\hat{b}_{j}^{2}+\hat{b}_{j}^{\dag 2})]
+J_{m}\hat{b}_{T}^{\dag}(\hat{b}_{L}+\hat{b}_{R})+h.c..
\end{eqnarray}
In this scheme, the coefficient $\Omega_{p}$ of this nonlinear interaction item is tunable utilizing the high precision electromagnetic technology, and the coupling strength $J_{m}$ between the mechanical modes can also be modulated via some electrical means, such as the capacitor method.

\subsection{The Hamiltonian of spin-cavity and cavity-cavity interactions}
As illustrated in Fig.~1 (a), a single NV center is set inside the central optical microcavity (mode $\hat{a}_{T}$), and this cavity mode will induce the transition between the states $|0\rangle$ and $|1\rangle$. Besides, this cavity also interacts with another two microcavity (modes $\hat{a}_{L}$ and $\hat{a}_{R}$) through exchanging photons.
Therefore, we can write the corresponding Hamiltonian in Schr\"{o}dinger picture (SP) as follow
\begin{eqnarray}
\hat{H}_{2}=\sum_{j=L,R,T}\omega_{c}\hat{a}_{j}^{\dag}\hat{a}_{j}+\omega_{A}\hat{\sigma}_{z}
+g\hat{a}_{T}^{\dag}\hat{\sigma}_{-}+J\hat{a}_{T}^{\dag}(\hat{a}_{L}+\hat{a}_{R})+h.c..
\end{eqnarray}
In which, $\omega_{c}$ is the fundamental frequency of these three identical cavities, $\omega_{A}$ stands for the energy-level transition frequency between the ground state $|m_{s}=0\rangle$ and excited state $|E_{y}\rangle$,
$g$ and $J$ mean the coupling strength for spin-cavity interaction and cavity-cavity interactions, respectively.

\subsection{The interactions between the mechanical modes and  cavity modes}
In this scheme, the three cavity modes $\hat{a}_{j}$ will also interact with the three mechanical modes $\hat{b}_{j}$, respectively. By setting the same coupling strength as
$g_{0}$ for simplicity and under the frame rotation with frequency $\omega_{p}$, we can express this type interaction with the form of the Hamiltonian,
\begin{eqnarray}
\hat{H}_{3}=\sum_{j=L,R,T}[-g_{0}\hat{a}_{j}^{\dag}\hat{a}_{j}(\hat{b}_{j}^{\dag}\exp{i\omega_{p}t}+h.c.)],
\end{eqnarray}
with the identical dispersive coupling strength $g_{0}$ between the cavity modes and the corresponding mechanical modes.

\subsection{The effective Hamiltonian derivation for this whole system}
Considering the Hamiltonian in Eq.~(1),
we can diagonalize the mechanical part of $\hat{H}_{1}$
by the unitary transformation $\hat{U}_{s}(r)=\exp[r(\hat{b}^{2}_{j}-\hat{b}^{\dagger2}_{j})/2]$,
where the squeezing parameter $r$ is defined via the relation $\tanh2r=\Omega_{p}/\Delta_{m}$.
In this squeezed frame, we can obtain the total Hamiltonian with the new expression
\begin{equation}
\hat{H}^{S}_{\text{Total}}=\hat{H}^{S}_{1}+\hat{H}^{S}_{2}+\hat{H}^{S}_{3},
\end{equation}
where
\begin{eqnarray}
\hat{H}_{1}^{S}=\sum_{j=L,R,T}\Delta^{S}_{m}\hat{b}_{j}^{S\dag}\hat{b}_{j}^{S}
+J_{m}^{S}\hat{b}_{T}^{S\dag}(\hat{b}_{L}^{S}+\hat{b}_{R}^{S})+h.c.,\notag
\end{eqnarray}
\begin{eqnarray}
\hat{H}_{2}^{S}=\hat{H}_{2},\notag
\end{eqnarray}
\begin{eqnarray}
\hat{H}_{3}^{S}=\sum_{j=L,R,T}[-g_{0}^{S}\hat{a}_{j}^{\dag}\hat{a}_{j}
(\hat{b}_{j}^{S\dag}+\hat{b}_{j}^{S})].\notag
\end{eqnarray}
In which, the relevant coefficients are respectively defined as:
\begin{eqnarray}
\alpha&=&\Omega_{p}/\Delta_{m},\notag\\
\Delta_{m}^{S}&=&\Delta_{m}(1-\alpha^{2})^{1/2},\notag\\
4r&=&\ln(1+\alpha)/(1-\alpha),\\
g_{0}^{S}&=&g_{0}e^{r}\cos \omega_{p}t,\notag\\
J_{m}^{S}&=&J_{m}e^{2r}/2.\notag
\end{eqnarray}

The mechanical part of this Hamiltonian can also be diagonalized by the canonical transformation
\begin{eqnarray}
\hat{b}_{T}^{S}&=&(\hat{b}_{+}-\hat{b}_{-})/\sqrt{2},\notag\\
\hat{b}_{L}^{S}&=&(\hat{b}_{+}+\hat{b}_{-}+\sqrt{2}\hat{b}_{0})/2,\\
\hat{b}_{R}^{S}&=&(\hat{b}_{+}+\hat{b}_{-}-\sqrt{2}\hat{b}_{0})/2,\notag
\end{eqnarray}
here, the modes $\hat{b}_{\pm,0}$ are the mechanical supermodes of the resonators.
Then we define the dimensionless position operators of these mechanical supermodes
$\hat{x}_{0}=\hat{b}_{0}^{\dag}+\hat{b}_{0}$ and $\hat{x}_{\pm}=\hat{b}_{\pm}^{\dag}+\hat{b}_{\pm}$.
In addition to the constraint condition $|\Delta_{m}^{S}\pm\sqrt{2}J_{m}^{S}|\gg |\omega_{A}-\omega_{c}|$,
we can discard modes $\hat{b}_{\pm}$ and only focus on the single mode $\hat{b}_{0}$ in this system.
Then the Hamiltonian $(\hat{H}_{1}^{S}+\hat{H}_{3}^{S})$ can be simplified as
\begin{eqnarray}
\hat{H}_{1}^{S}+\hat{H}_{3}^{S}&\simeq&\Delta_{m}^{S}\hat{b}_{0}^{\dag}\hat{b}_{0}
+\frac{g_{0}^{S}}{\sqrt{2}}\hat{x}_{0}(\hat{a}_{R}^{\dag}\hat{a}_{R}-\hat{a}_{L}^{\dag}\hat{a}_{L}).
\end{eqnarray}

Applying an unitary transformation with the definition
$\hat{U}=\exp[-i\omega_{c}(\hat{a}_{L}^{\dag}\hat{a}_{L}
+\hat{a}_{T}^{\dag}\hat{a}_{T}+\hat{a}_{R}^{\dag}\hat{a}_{R}
+\hat{\sigma}_{z})t]$ to the Hamiltonian $\hat{H}_{\text{Total}}^{S}$,
we can obtain
\begin{eqnarray}
\hat{H}_{\text{Total}}^{S}=\Delta\hat{\sigma}_{z}+\Delta_{m}^{S}\hat{b}_{0}^{\dag}\hat{b}_{0}+\hat{\Theta}(\hat{a}_{R}^{\dag}\hat{a}_{R}-\hat{a}_{L}^{\dag}\hat{a}_{L})
+g\hat{a}_{T}^{\dag}\hat{\sigma}_{-}+J\hat{a}_{T}^{\dag}(\hat{a}_{L}+\hat{a}_{R})+h.c.,
\end{eqnarray}
where $\Delta=\omega_{A}-\omega_{c}$ and $\hat{\Theta}=g_{0s}\hat{x}_{0}/\sqrt{2}$.

Then we can also diagonalize the cavity-mode part in the Hamiltonian $\hat{H}_{\text{Total}}^{S}$ by introducing another canonical transformation
\begin{eqnarray}
\hat{a}_{0}&=&-r_{1}\hat{a}_{L}+r_{2}\hat{a}_{T}+r_{1}\hat{a}_{R},\notag\\
\hat{a}_{+}&=&r_{3}\hat{a}_{L}-r_{1}\hat{a}_{T}+r_{4}\hat{a}_{R},\\
\hat{a}_{-}&=&r_{4}\hat{a}_{L}+r_{1}\hat{a}_{T}+r_{3}\hat{a}_{R},\notag
\end{eqnarray}
where $r_{1}=J/E$, $r_{2}=\hat{\Theta}/E$, $r_{3}(\hat{\Theta},J)=r_{4}(-\hat{\Theta},J)$, and $E=\sqrt{2J^{2}+\hat{\Theta}^{2}}$.
Then the Hamiltonian $\hat{H}_{\text{Total}}^{S}$ can also be further simplified as
\begin{eqnarray}
\hat{H}_{\text{Total}}^{S}=\Delta\hat{\sigma}_{z}+\Delta_{m}^{S}\hat{b}_{0}^{\dag}\hat{b}_{0}
+E(\hat{a}_{+}^{\dag}\hat{a}_{+}-\hat{a}_{-}^{\dag}\hat{a}_{-})
+g r_{2}\hat{a}_{0}^{\dag}\hat{\sigma}_{-}+gr_{1}(\hat{a}_{+}^{\dag}-\hat{a}_{-}^{\dag})\hat{\sigma}_{-}+h.c..
\end{eqnarray}
When the spectral separation
between the supermodes is much larger than the mechanical frequency ($E\gg\Delta_{m}^{S}$),
we can neglect the effect of the terms of supermodes $\hat{a}_{\pm}$ and
get the effective Hamiltonian by utilizing the approximate relation $r_{2}=\hat{\Theta}/E\approx g_{0s}\hat{x}_{0}/2J$,
\begin{eqnarray}
\hat{H}_{\text{eff}}\approx\Delta\hat{\sigma}_{z}+\Delta_{m}^{S}\hat{b}_{0}^{\dag}\hat{b}_{0}
+\frac{gg_{0}^{S}}{2J}(\hat{b}_{0}+\hat{b}_{0}^{\dag})(\hat{a}_{0}^{\dag}\hat{\sigma}_{-}+h.c.).
\end{eqnarray}

\section*{Funding}
Natural National Science Foundation (NSFC) (11774285, 12047524, 11774282, 11504102);
China Postdoctoral Science Foundation (2021M691150);
Natural Science Foundation of Hubei Province (2020CFB748, 2019CFB788);
Research Project of Hubei Education Department (B2020079,D20201803);
Program for Science and Technology Innovation Team in Colleges of Hubei Province (T2021012);
Doctoral Scientific Research Foundation of Hubei University of Automotive Technology (HUAT) (BK201906, BK202106, BK202008, BK201804);
Open Fund of HUAT (QCCLSZK2021A07);
Foundation of Discipline Innovation Team of HUAT.
\section*{Acknowledgments}
Y. Z. thanks Xin-You L\"{u} and Peng-Bo Li for valuable discussions. Part of the simulations are coded
in $\textsc{python}$ using the $\textsc{qutip}$ library \cite{Johansson2012QuTiP, Johansson2013QuTiP}.

\section*{Disclosures}
The authors declare that there are no conflicts of interest related to this article.


\bibstyle{acm}
\bibliography{ref}






\end{document}